\documentclass{appolb}
\usepackage{graphicx}
\usepackage{authblk}
\usepackage[utf8]{inputenc} 
\usepackage{fancyhdr}
\usepackage{indentfirst}
\usepackage{mathtools}
\usepackage{amsmath}
\usepackage{cite}

\begin{document}
\title{Beam profile investigation of the new collimator system for the J-PET detector}
\author{E.~Kubicz$^{1}$, M.~Silarski$^{1}$,  A.~Wieczorek$^{1,}$ $^{2}$, D.~Alfs$^{1}$, T.~Bednarski$^{1}$, P.~Bia{\l}as$^{1}$, E.~Czerwi{\'n}ski$^{1}$, A.~Gajos$^{1}$, B.~Głowacz$^{1}$, B.~Jasi{\'n}ska$^{3}$, D.~Kami{\'n}ska$^{1}$, G.~Korcyl$^{1}$, P.~Kowalski$^{4}$, T.~Kozik$^{1}$, W.~Krzemie{\'n}$^{5}$, M.~Mohammed$^{1}$, I.~Moskal$^{1}$, S.~Nied{\'z}wiecki$^{1}$, M.~Pawlik-Nied{\'z}wiecka$^{1}$, L.~Raczy{\'n}ski$^{5}$, Z.~Rudy$^{1}$, A.~Strzelecki$^{1}$, W.~Wi{\'s}licki$^{5}$, M.~Zieli{\'n}ski$^{1}$, B.~Zgardzi{\'n}ska$^{3}$, P.~Moskal$^{1}$}
\maketitle
$^{1}$ Faculty of Physics, Astronomy and Applied Computer Science, Jagiellonian University, 
S.~{\L}ojasiewicza 11, 30-348 Krak{\'o}w, Poland

       $^{2}$ Institute of Metallurgy and Materials Science of Polish Academy of Sciences, 
W.~Reymonta 25, 30-059 Krak{\'o}w, Poland
       
             $^{3}$ Department of Nuclear Methods, Institute of Physics, Maria Curie-Sklodowska University, Pl. M. Curie-Sklodowskiej 1, 20-031 Lublin, Poland
       
       $^{4}$ {\'S}wierk Computing Centre, National Centre for Nuclear Research, 
A.~Soltana 7, 05-400 Otwock-{\'S}wierk, Poland
             
      $^{5}$ High Energy Physics Division, National Centre for Nuclear Research,
A.~ Soltana 7, 05-400 Otwock-{\'S}wierk, Poland

\begin{abstract}
  Jagiellonian Positron Emission Tomograph (J-PET) is a multi-purpose detector which will be used for search for discrete symmetries violations in the decays of positronium atoms and for investigations with positronium atoms in life-sciences and medical diagnostics. In this article we present three methods for determination of the beam profile of collimated annihilation gamma quanta. Precise monitoring of this profile is essential for time and energy calibration of the J-PET detector and for the determination of the library of model signals used in the hit-time and hit-position reconstruction. We have we have shown that usage of two lead bricks with dimensions of 5x10x20 cm$ ^{3} $ enables to form a beam of annihilation quanta with Gaussian profile characterized by 1~mm FWHM. Determination of this characteristic is essential for designing and construction the collimator system for the 24-module J-PET prototype. Simulations of the beam profile for different collimator dimensions were performed. This allowed us to choose optimal collimation system in terms of the beam profile parameters, dimensions and weight of the collimator taking into account the design of the 24 module J-PET detector.

\end{abstract}

\newpage
\section{Introduction}

\indent Positron emission tomography (PET) is a well recognized diagnostic method enabling imaging of chosen substance's metabolism in a living organism. The state-of-the-art commercial PET scanners are based on inorganic scintillators as radiation detectors \cite{1,2}. This PET technology is expensive \cite{3,4} and therefore there are attempts to find a new, more affordable solutions as e.g. described in ref. \cite{5,6}. 

\indent The J-PET group is developing a cost-effective whole-body positron emission tomography scanner based on plastic scintillators \cite{7,8,9,10,11,12,neha,13,14,15}. The scanner, referred to as J-PET (Jagiellonian Positron Emission Tomograph) is built out of axially arranged plastic scintillator stripes, read out at both sides by photomultipliers \cite{9}. Signals from photomultipliers are sampled in the voltage domain by the dedicated front-end electronics \cite{15}. Based on the difference and the average of these signals it is possible to reconstruct both position and time of the gamma quantum interaction with scintillator, respectively. Currently three reconstruction methods were developed by the J-PET group. One approach is to calculate the Mahalanobis distances of the measured signal (represented by the multivariate vector) from the vectors based on the mean of the signals in data sets determined for known positions \cite{11,neha}. In the second method the hit time and hit position are reconstructed based on the comparison of the measured signal probed in the voltage or time domains with the synchronized model signals \cite{10}. The third option is to train data set from a multivariate normal distribution based on the Tikhonov regularization, where representation of signals was provided by the Principal Component Analysis decomposition \cite{12,13}. All these three methods required data sets collected for know position of gamma interaction with scintillator. Annihilation quanta were collimated with the usage of lead bricks with 1.3~mm slit. However, due to the finite slit length and the distance between collimator and the scintillator stripe, the precise determination of the beam profile was performed. For the first tests we have built two different collimation systems shown on Fig.~1~and~2. These setups consisted of Hammamatsu R4998 photomultipliers \cite{16}, reference detector, lead bricks and BC-420 scintillators manufactured by Saint-Gobain \cite{17}. Signals from photomultiplier were sampled in time domain by the LeCroy SDA6000A digital oscilloscope. 

\begin{figure}[h!]
\begin{center}
\includegraphics [scale=0.4] {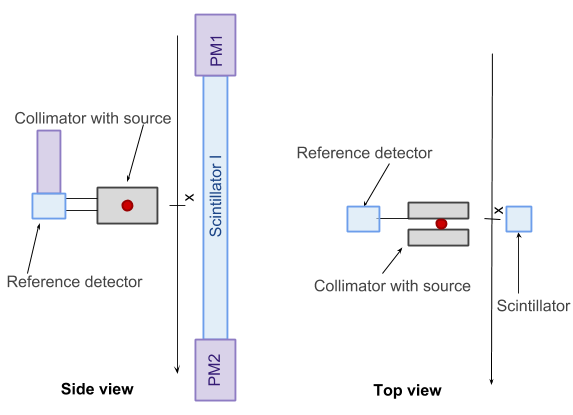}
\caption{Scheme of the setup used in method I. $^{68}$Ge source was placed in the collimator built out of two lead bricks each with dimensions of 55x100x200 mm$ ^{3} $ (slit 1.3~mm). 
M(x) was measured as a number of coincident signals in the reference and scintillator detectors. Figure is not to scaled.}
\end{center}
\end{figure}

First setup consisted of a scintillator with two photomultipliers, one for the each scintillator end, and a reference detector fixed mechanically to the collimator. Scintillator was aligned perpendicularly to the axis along which collimator with the $^{68}$Ge source was moved (Fig.~1). For the second setup two scintillator stripes with photomultipliers attached to each end were placed parallel to each other. Collimator (built of lead cylinders) with $^{22}$Na  source  was placed between them and was moved along the axis parallel to the two scintillator stripes. Additionally, one or two lead bricks were placed between the source and one scintillator (Fig.~2). We have also developed three different methods to determine the profile of the  annihilation gamma quanta beam. 

\begin{figure}[h!]
\begin{center}
\includegraphics [scale=0.49] {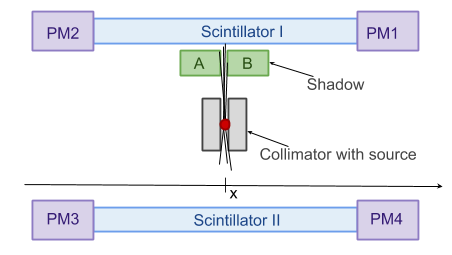}
\caption{Experimental setup for method II and III of the beam profile measurement. $ x $ - position of the collimator with source. Between the collimator with $ ^{22}$Na source (slit 1.3~mm) and one of the scintillators with size of 5x19x300~mm$ ^{3} $ two lead bricks with dimensions of 50x70x70 mm$ ^{3} $  ("shadow") with slit of 0.3~mm were placed. M(x) was measured as a number of coincident signals in both scintillator detectors. Figure is not to scaled. }
\end{center}
\end{figure}

\indent  The first prototype of J-PET which have a downscaled geometry of tomographs used in hospitals  is build out of 24-modules (scintillator stripe with pair of photomultipliers) (Fig.~3). This required a design of a new collimation system for the purposes of signal database measurements and any other studies of the detector response requiring well collimated beam \cite{11}. In order to find a proper dimensions of the lead cylinders providing the best beam collimation and low total weight of the collimator we have made simple calculations neglecting gamma quanta scattering. These calculations give
also the first approximations of expected beam profiles for the new collimator.

\begin{figure}[h!]
\begin{center}
\includegraphics [scale=0.35] {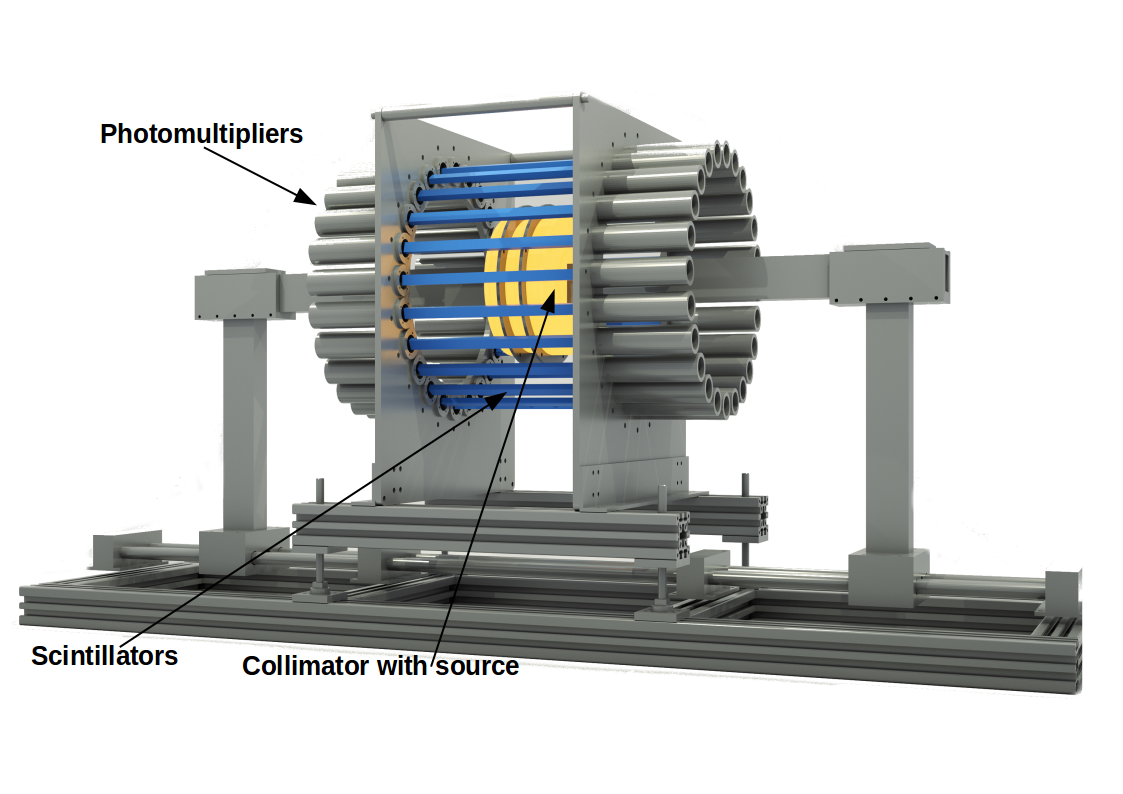}
\caption{Drawing of the  first J-PET detector composed of 24 modules together with a mechanical structure for holding the cylindrical collimator.}
\end{center}
\end{figure}

\newpage
\section{Measurements of the beam profile for double-stripe J-PET prototype}

\indent For better control over the systematic uncertainties of the beam profile determination we have developed three quantitatively different methods (described in sections 3 - 5). Each method required specific configuration of an experimental setup presented in Fig.~1 and Fig.~2. First method allows  to measure the cumulative function of the beam profile without modifying the beam itself. In method II  part of the beam is being absorbed by lead brick placed between source and scintillator. Method III allows to scan beam sections analogously as it is in laser optics and measure beam profile directly. \\
\indent In order to determine the beam profile $ h(x) $ the scintillation detector was irradiated by the tested beam which can be moved along the x-axis. The measured number of events $ M(x) $ as a function of x can be expressed as a convolution of the beam profile and the detector acceptance function $ g(x) $:

\begin{center}
\begin{equation}
M(x)= h(x)\ast g(x)= \int\limits^{+\infty}_{-\infty}h(x-x')g(x')dx' .
\end{equation}
\end{center}

\indent In order to select annihilation quanta a coincident registration of signals in both detectors was required. In the first method the collimator and the reference detector are fixed together and are moved with respect to the scintillator. In the second method an additional lead brick (A) is used as a "shadow" to absorb a part of the beam, and in the third method two lead bricks (A,B) form a narrow slit allowing to make a direct scan of a beam profile (Fig.~2). 

\section{Method I}

\indent The experimental setup used for this method is shown in Fig.~1. The function of the geometrical acceptance of the detector can be approximated by:

\begin{center}
\begin{equation}
g(x) = \left \{ {{1\ \ \text{if }\ \  x\in[x_a, x_b] } \atop{0\ \ \text{if } \  x\not\in [x_a, x_b] }} \right.
\end{equation}
\end{center}

\noindent where: $x_{a}$ and  $x_{b}$  denote positions of the beginning and the end of the scintillator.
	This implies that only these gamma quanta which are within the section [$x_{a}$, $x_{b}$] can generate a signal. A beam profile $ h(x) $ can be obtained by applying equation (2) into equation (1) and subsequent differentiation of equation (1):
	
\begin{center}
\begin{equation}
\frac{d}{dx} M(x) = h(x-x_b) - h(x-x_a) 
\end{equation}
\end{center}

\indent  In the left panel of Fig. 4 distribution M(x) obtained in the measurement is shown. In order to extract the shape of the beam profile the numerical derivative of $ M(x) $ was calculated as follows:

\begin{center}
\begin{equation}
h(x) = \frac{dM}{dx} = \frac{N_{2}-N_{1}}{x_{2}-x_{1}}
\end{equation}
\end{center}

\noindent where: $N_{1}$, $N_{2}$ denote number of counts for measurements at $x_{1}$, $x_{2}$, with $x = (x_{2}$ - $x_{1})/2$. The result is shown in the right panel of Fig.~4. 

\begin{figure}[h!]
\begin{center}
\includegraphics [scale=0.058] {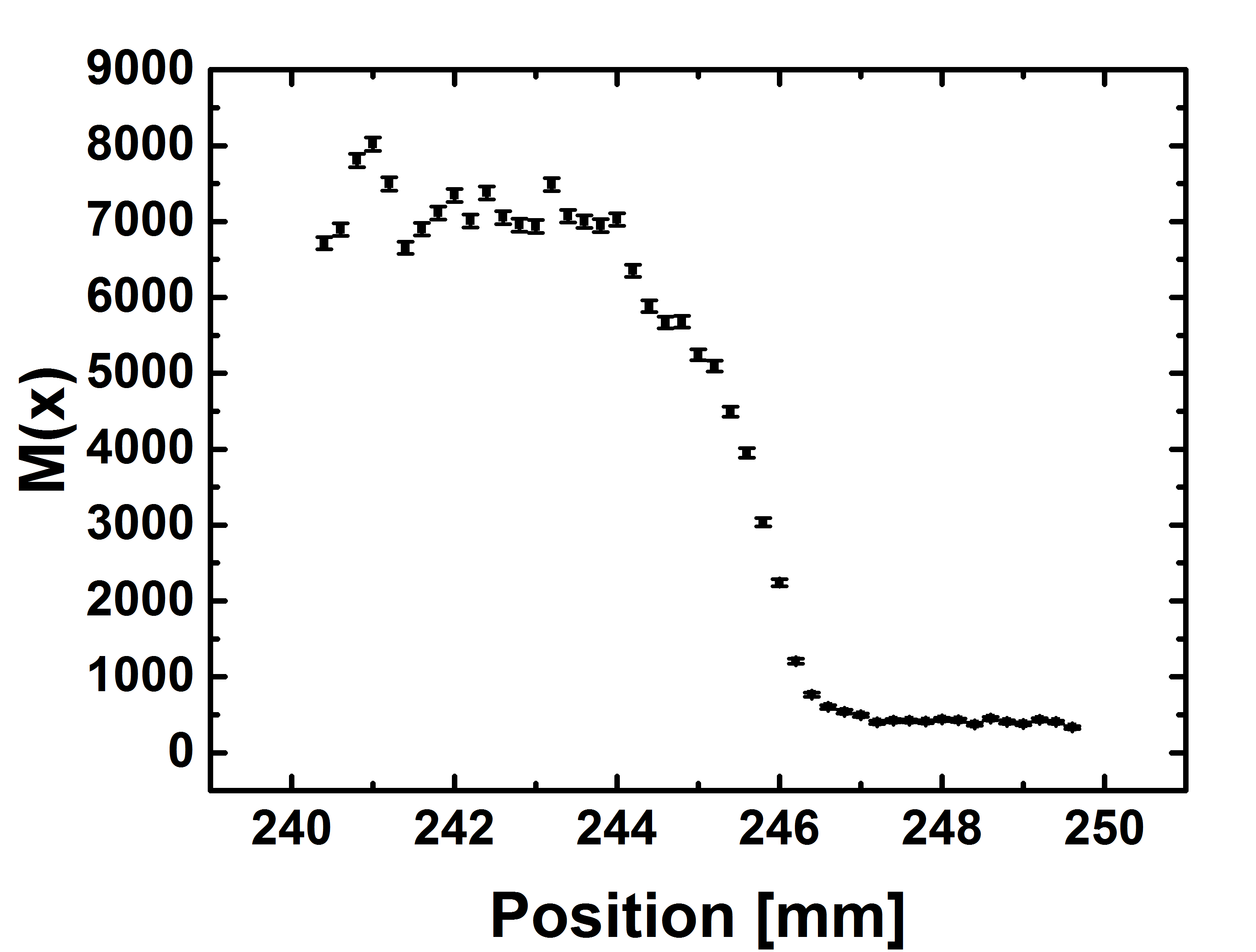}
\includegraphics [scale=0.058] {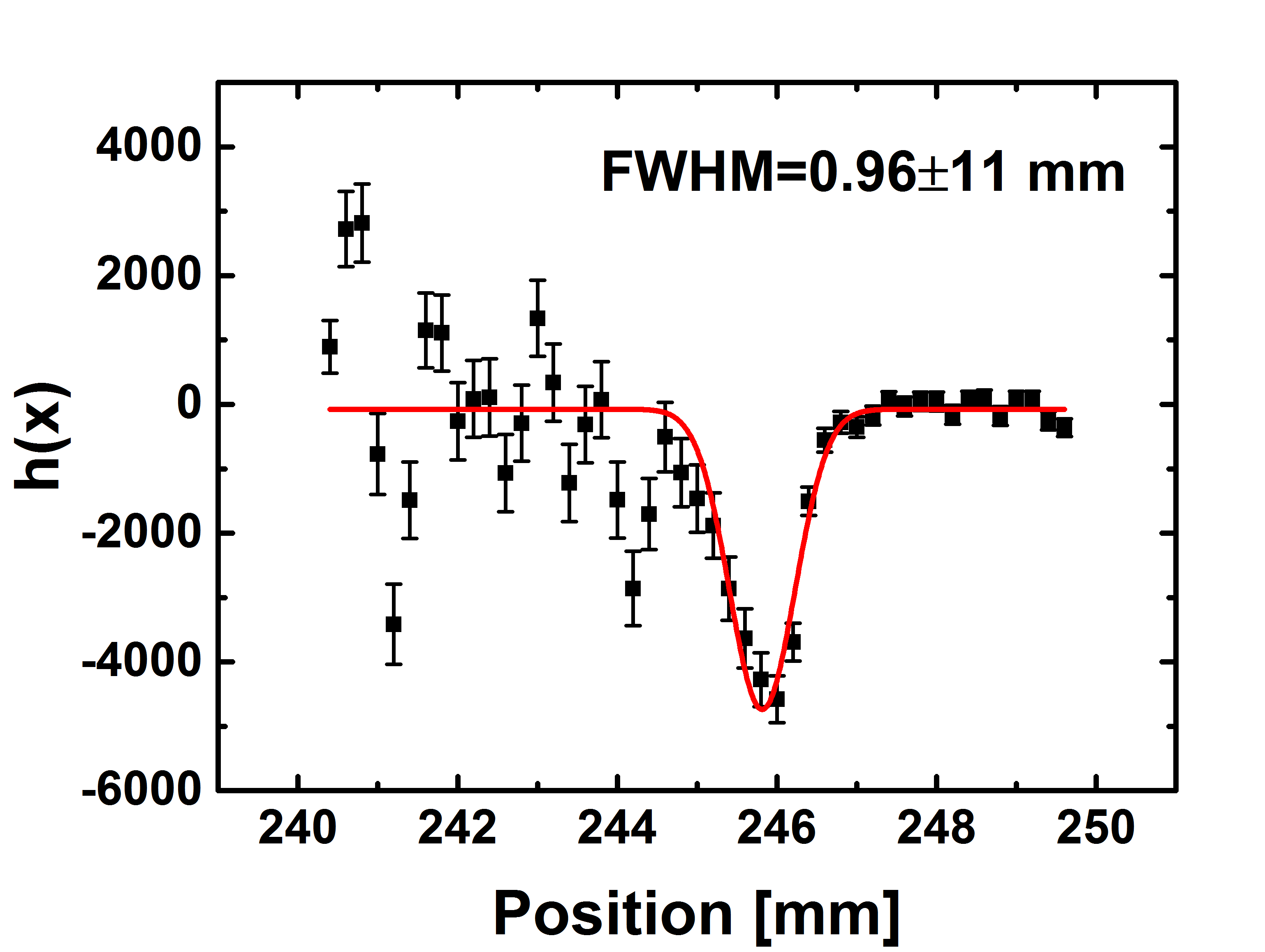}
\caption{(left)  Number of events measured per 4 minutes as a function of the position of the source. (right) Derived beam profile $ h(x) $.  Superimposed red line indicates result of the Gauss function fit to the data. The obtained  FWHM of the beam profile is equal to 0.96(11)~mm.  Absolute values on the horizontal axis are given in the scintillator reference frame.}
\end{center}
\end{figure}

\section{Method II}

\indent Fig.~2 shows experimental setup used for methods II and III. Measurements were done by moving the collimator with a dedicated mechanical system along the x-axis with steps of 0.3~mm.

\indent	In case of method II only one lead block - $ A $ - absorbing gamma quanta was used. Since the block size is much larger than the beam profile we assume that the lead block absorbs gamma quanta in the range from $x_{0}$ to $ + \infty $, where $x_{0}$ denotes beginning of the block. In this case the function of the geometrical detector acceptance can be approximated by:

\begin{center}
\begin{equation}
g(x) = \left \{ {{0\ \ \text{if }\ \  x\in(x_0, +\infty) } \atop{1\ \ \text{if } \  x\in (-\infty, x_{0}] }} \right.
\end{equation}
\end{center}

and hence the derivation of $ M(x) $ yields:

\begin{center}
\begin{equation}
M(x) = h(x)\ast g(x) = \int\limits^{x_0}_{-\infty} h(x-x')dx'
\end{equation}
\end{center}

In Fig.~5 an exemplary distribution of $ M(x) $ and beam profile obtained in the measurement are shown.

\begin{figure}[h!]
\begin{center}
\includegraphics [scale=0.057] {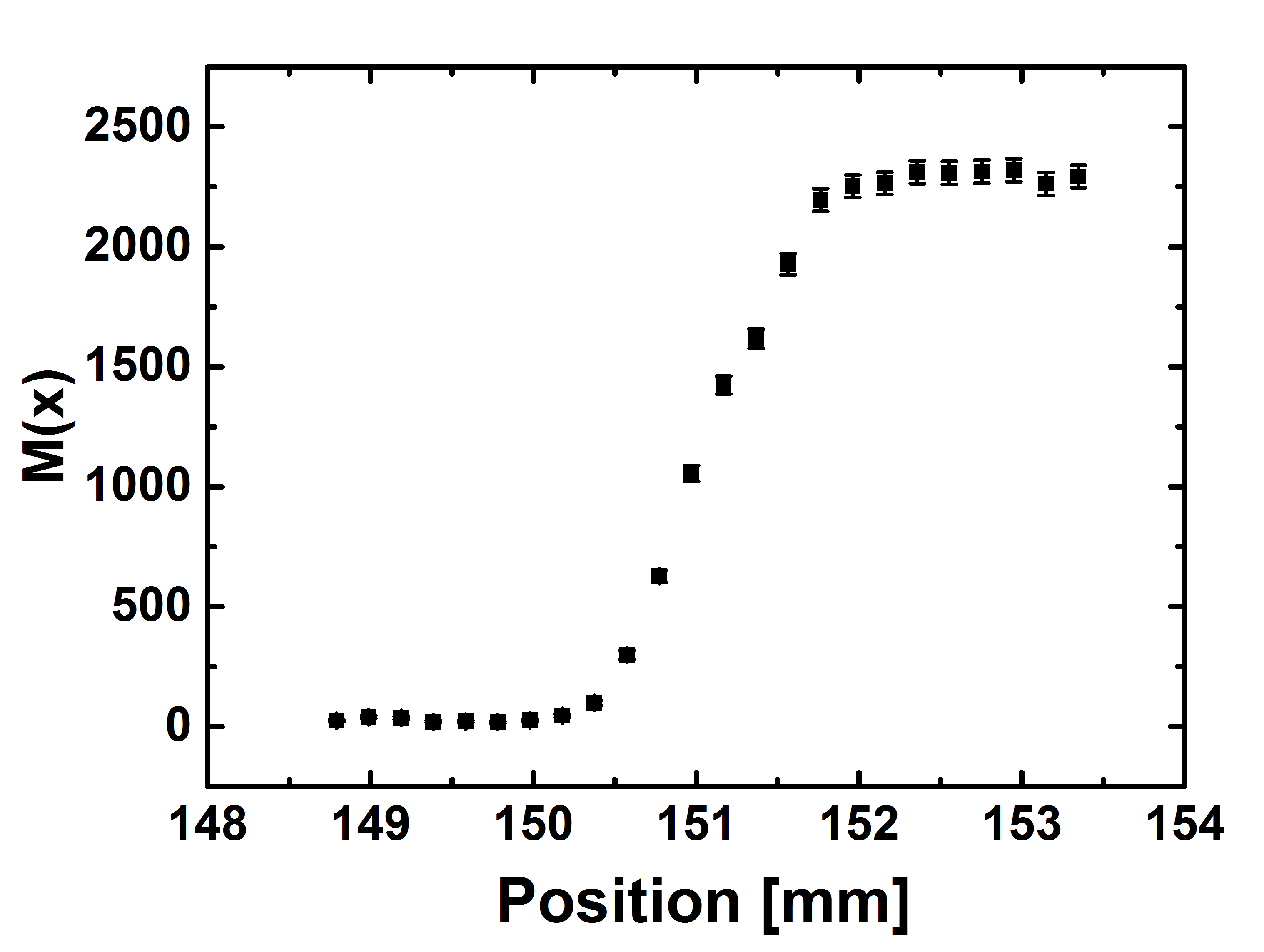}
\includegraphics [scale=0.057] {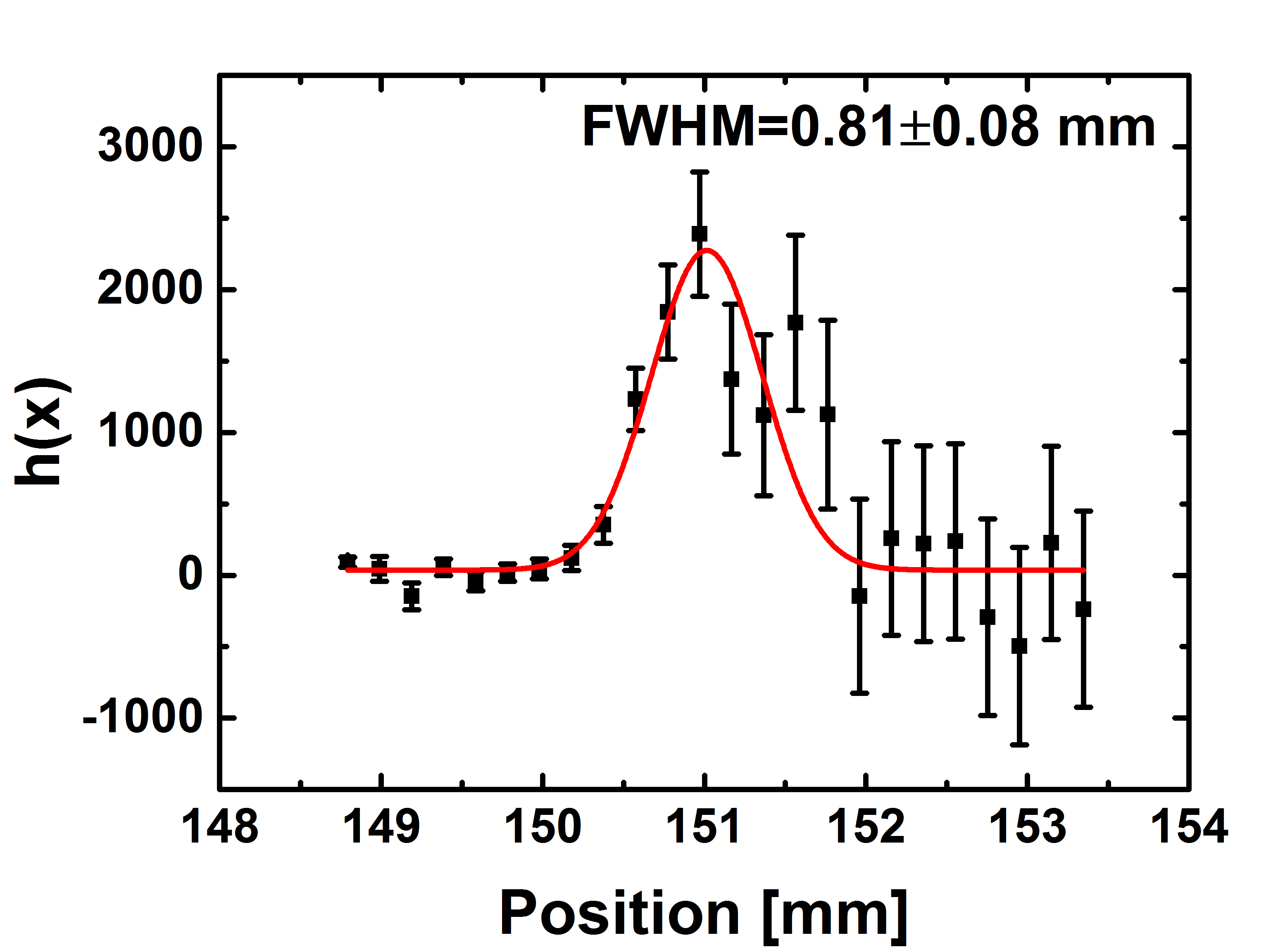}
\caption{(left) Number of coincidences as a function of the relative position between the collimator and  bar A of the shadow. The slit width of the collimator was equal to 1.3 mm. Scintillators were aligned horizontally. (right) Derived beam profile $ h(x) $. FWHM derived from the fit Gauss function amounts to 0.90(10) mm.}
\end{center}
\end{figure}

\section{Method III}

\indent In this part of experiment two lead bricks were placed with 0.3 mm wide slit parallel to the collimator slit (Fig.~2). By moving a collimated beam along the x-axis we have received beam profile directly from the measurement of $ M(x) $, without any additional calculation. Below in Fig.~6 results from measurement with two lead bars as a shadow are shown. The reason for the asymmetric shape of the beam profile is not perfectly smooth bar's surfaces.

\begin{figure}[h!]
\begin{center}
\includegraphics [scale=0.06] {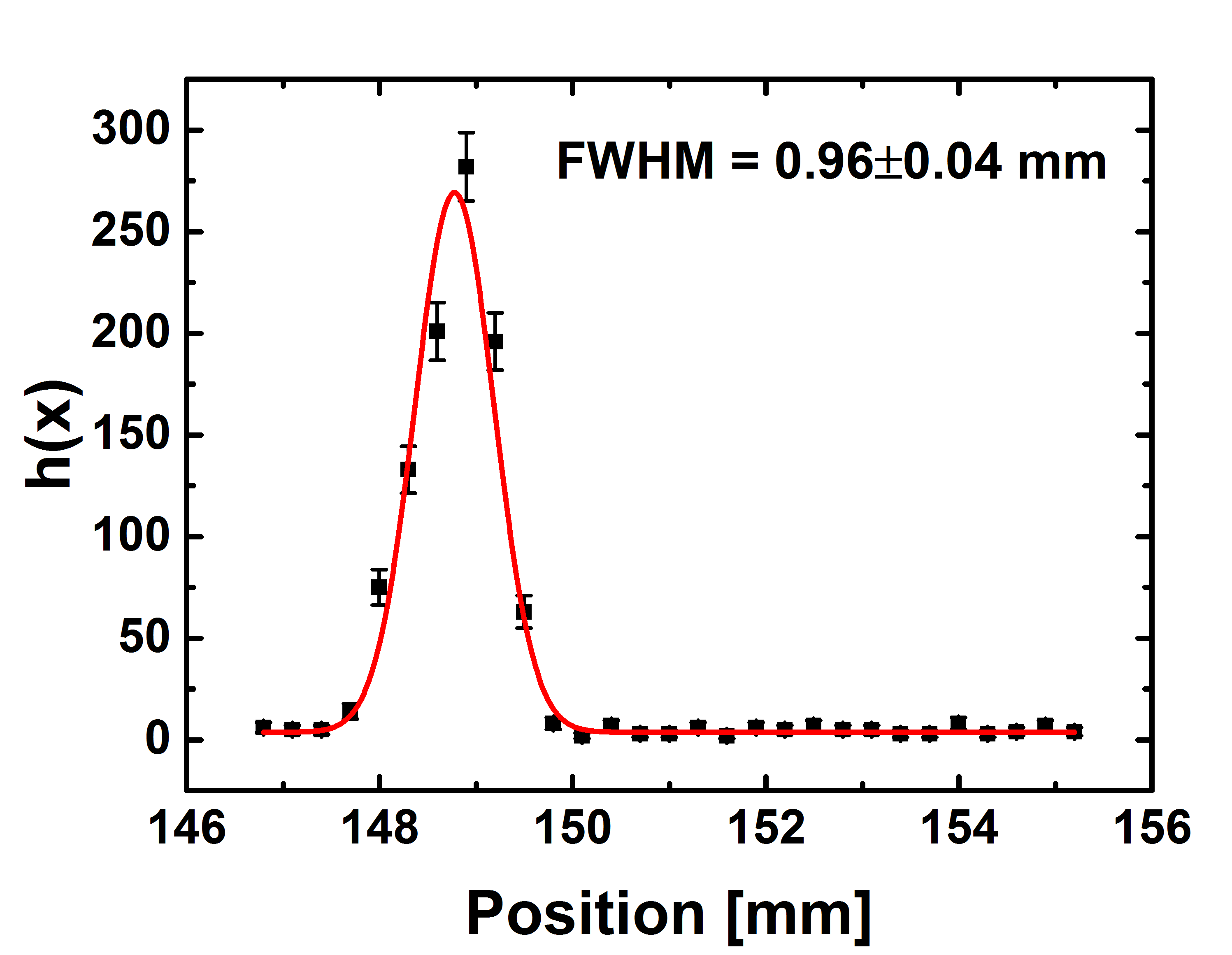}
\caption{Number of events measured in 15 min intervals as function of the position of the source. The FWHM derived from the fit Gauss function amounts to 0.96(04) mm.}
\end{center}
\end{figure}

As one can see in Tab.~1, obtained values of the width of the beam profile are compatible within the statistical uncertainties for all presented methods and they are equal to about ~1~mm. The reason for big statistical errors that occur in methods I and II are mathematical transformations, which allow to extract beam profile form $ M(x) $ function.  Therefore to reduce impact of these calculations method III can be used as a model for the beam profile measurements. One must take into account that it rejects all gamma quanta which do not travel perpendicular to the scintillator, however as it is shown in this paper results for the third method are the same as for two previous one but with much smaller statistical errors. As a result influence of 'shadow' bars on the beam can be neglected.

\begin{table}[h!]
\centering
\caption{Values of FWHM as a beam profile width from Gaussian fit.}
\vspace{0.5cm}
\begin{tabular}{|c|c|}
\hline
\textbf{Measurement} & \textbf{FWHM [mm]}   \\\hline
Method I & 0.96(11)   \\\hline
Method II & 0.90(10)   \\\hline
Method III & 0.96(04)  \\\hline
\end{tabular}
\end{table}

\section{Simulations of beam profile for different collimator systems for 24 module J-PET prototype}

\indent The new collimator for the 24-module barrel tomograph will take advantage of the symmetry of the detector, and thus it consist of two lead cylinders connected with screws (size of which can be neglected regarding the scattering effects in further considerations) placed in a way such, that the beam of gamma quanta will irradiate all the detection modules at the same time. Moreover, it should provide a precise position determination of the beam. In that case the main difficulty which we have to deal with is the weight of the cylinders. Due to high density of lead arms supporting the collimator are exposed to big bending moments which may destroy the symmetry of the whole system and prevent precise measurements. Due to cylindrical symmetry of both the detector and collimator, the whole problem can be reduced to two dimensions. The schematic view of the assumed geometry is shown in Fig.~7. The two lead cylinders with radius $ r $ and height $ h $ are distant from each other by $ s $ (slit). The distance between scintillators amounts to $ 2R $. We assume the point-like $ ^{22}$Na source emitting gamma quanta isotropically in every direction. As it was already mentioned the scattering of gamma quanta is neglected assuming only one act of interaction with material of the collimator.

\begin{figure}[h!]
\begin{center}
\includegraphics [scale=0.5] {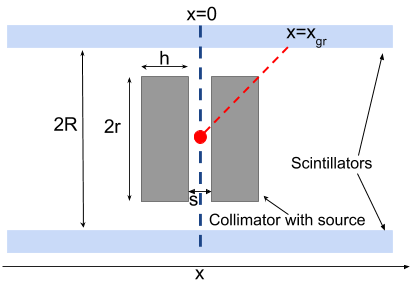}
\caption{The scheme of the system assumed in simulations. The source is placed in the geometric center of the collimator. Taking advantage of the cylindrical geometry we can consider only one cross-section in the $ x-y $ plane.}
\end{center}
\end{figure}

\indent The beam of gamma quanta passing through the collimator material is attenuated according to the equation:

\begin{center}
\begin{equation}
N(z) = N_{0} e^{-\mu z}
\end{equation}
\end{center}

\noindent where: $ N $ - the number of gamma quanta which passed through depends only on the length of the path $ z $ in the material (if it is homogeneous).\\
The attenuation length  $ \mu $  depends on the material and the energy of the photon:

\begin{center}
\begin{equation}
\mu = k \sigma = \frac{N_{A} \rho}{m_{mol}} \sigma_{Pb}
\end{equation}
\end{center}

\noindent where: $ N_{A}$ is the Avogadro number, $ \rho $ and $ m_{mol} $ denote the density and molar mass of the material, respectively. $ \sigma_{Pb} $  stands for the total cross sections of gamma quanta interaction with lead. Since we are dealing with the emitted gamma quanta of energy $ E_{\gamma} = 0.511 MeV $ passing through lead the value of $ \mu $ amounts to $ 1.73 \frac{g}{mol*cm} $ (where $ \sigma_{Pb} = 5.223*10^{-23} cm^{2} = 52.23~b $). 

\indent To determine the $ z(x) $ dependence for annihilation quanta we have par\-ametrized the path length $ z $ with the position along the scintillator $ x $. Using the symmetry (since we assume the source position exactly in the geometric center) it is enough to consider only values of $x \geq 0 $ . The $ z(x) $ dependence can be then expressed as:

\begin{center}
\begin{equation}
z(x) = \left \{ {{\left( r - \frac{sR}{2x} \right) \sqrt{\frac{x^{2}}{R^{2}} +1},   x > x_{gr} } \atop{\frac{dR}{x} \sqrt{\frac{x^{2}}{R^{2}} +1}, x \leq x_{gr} }} \right.
\end{equation}
\end{center}

\noindent where: $ x_{gr}= \left(  d + \frac{s}{2} \right)  \frac{R}{r}$ is the position for which $ z $ is maximum (see Fig.~7). This parametrization allows to calculate the probability of annihilation gamma quantum passing through the collimator and as a consequence, the beam profile. These calculations were done for different dimensions of the collimator, as one can see in the Tab.~2, assuming $ R=175~mm $.

\begin{table}[h!]
\centering
\caption{The summary of results obtained for $ s~=~1~mm $ for the beam formed by annihilation gamma quanta passing through the collimator of different assumed geometrical dimensions.}
\vspace{0.5cm}
\begin{tabular}{|c|c|c|c|}
\hline
 \textbf{r [mm]} & \textbf{h [mm]} & \textbf{FWHM/2 [mm]} & \textbf{Weight [kg]}    \\\hline
 100 & 30 & 0.91 & 10.7   \\\hline
 100 & 50 & 0.91 & 17.9   \\\hline
 50 & 100 & 1.90 & 8.9   \\\hline
 50 & 50  & 1.90 & 4.5   \\\hline
 100 & 10 & 0.92 & 3.6   \\\hline
 80 & 50 & 1.16 & 11.5   \\\hline
 80 & 30 & 1.16 & 6.9   \\\hline
 90 & 30 & 1.10 & 8.7   \\\hline
 70 & 30 & 1.34 & 5.3   \\\hline
 60 & 50 & 1.56 & 6.4  \\\hline
\end{tabular}
\end{table}

\newpage
In Fig.~8~and~9 results of beam profile calculation for two chosen collimator systems are shown.

\begin{figure}[h!]
\begin{center}
\includegraphics [scale=0.057] {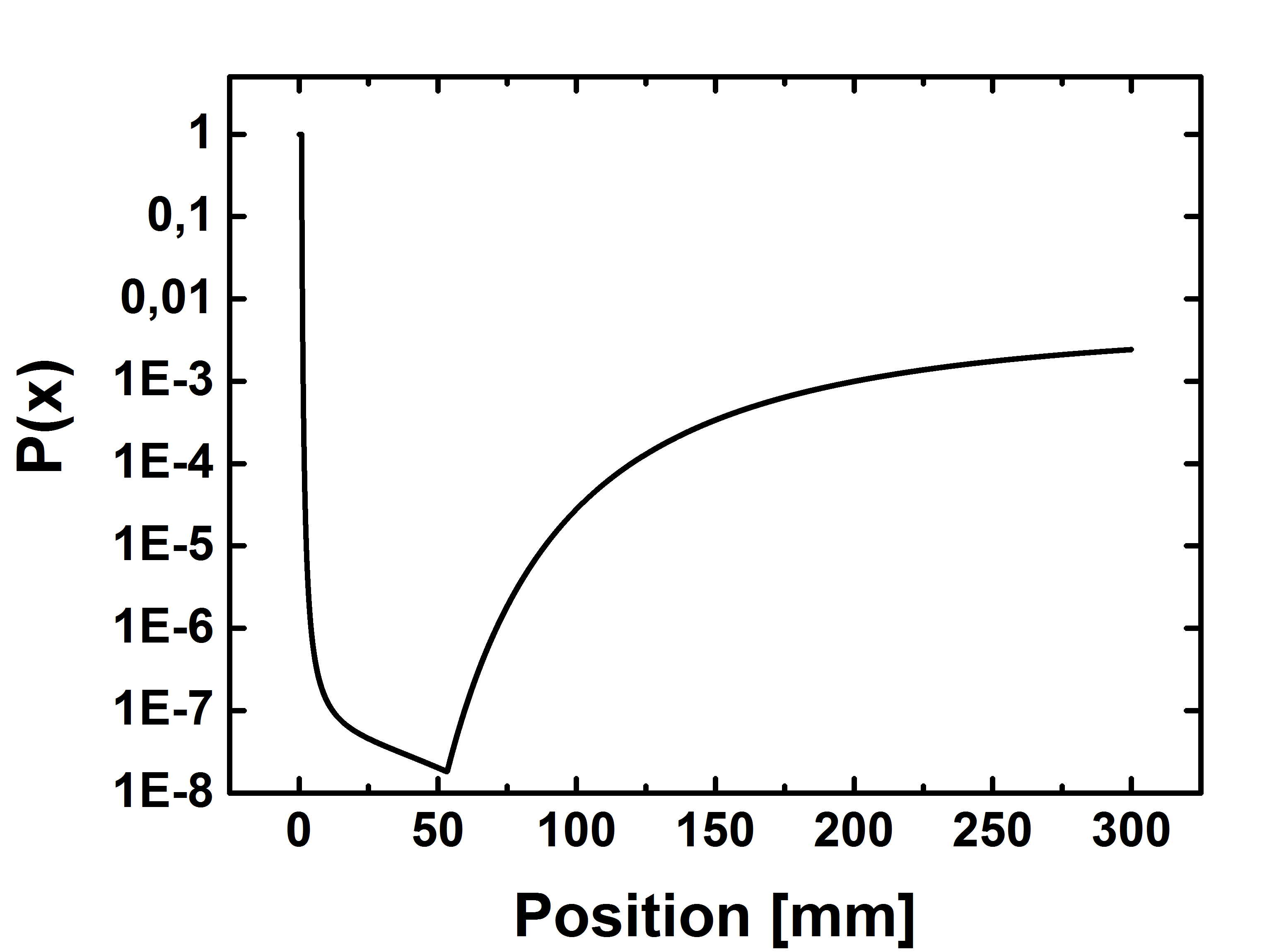}
\includegraphics [scale=0.057] {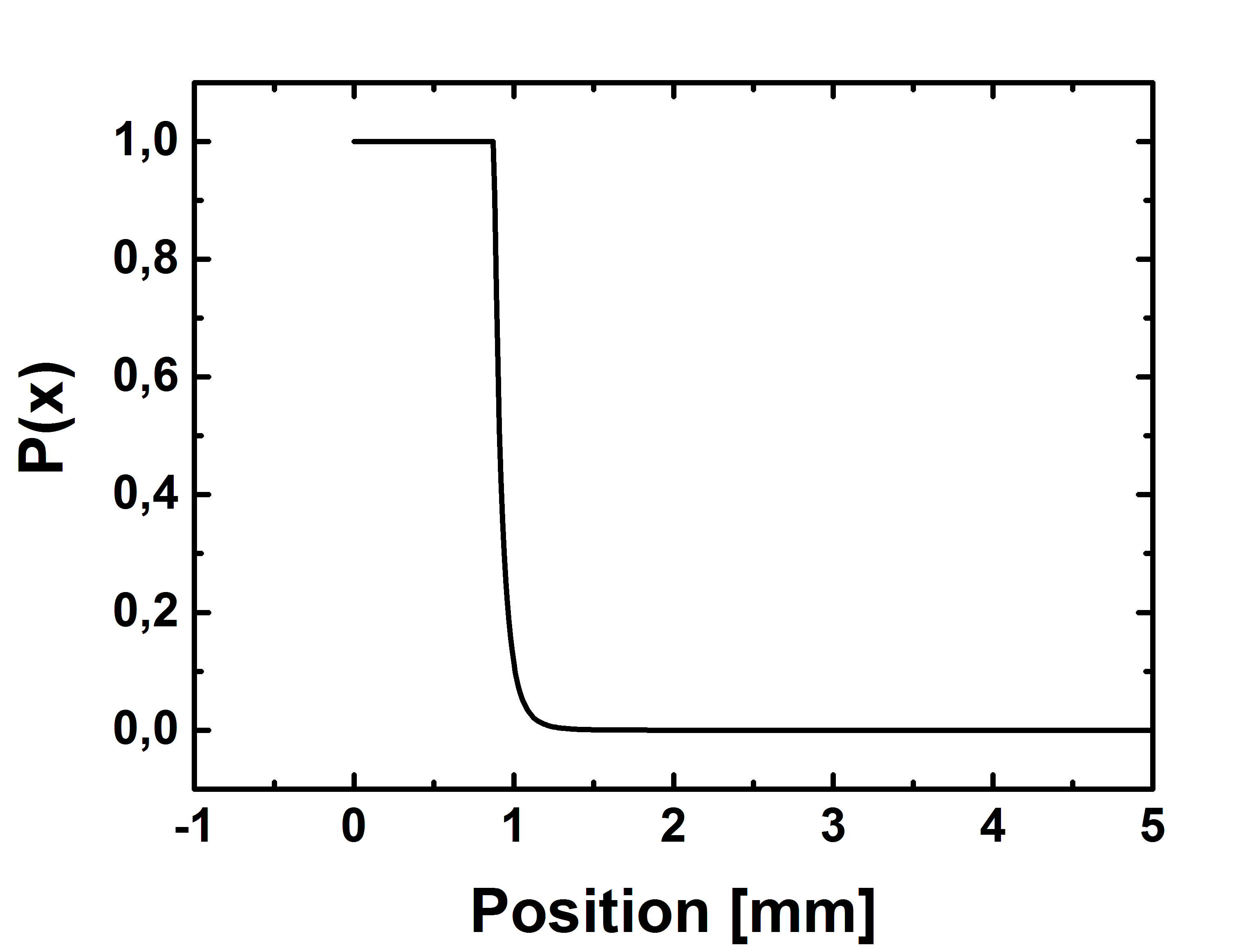}
\caption{(left)The probability of annihilation gamma quantum to pass through the collimator as a function of the position along the stripe ($x>0$) for $ s~=~1~mm $, $ r~=~100~mm $ and $ h~=~30~mm $ in logarithmic scale. (right) Zoom of the distribution for $0<x<5~mm$ in linear scale. The measure of profile of the beam was defined as the $ x $ position for which $ P(x)$ $ \simeq 0.5$ multiplied by two.}
\end{center}
\end{figure}

\begin{figure}[h!]
\begin{center}
\includegraphics [scale=0.058] {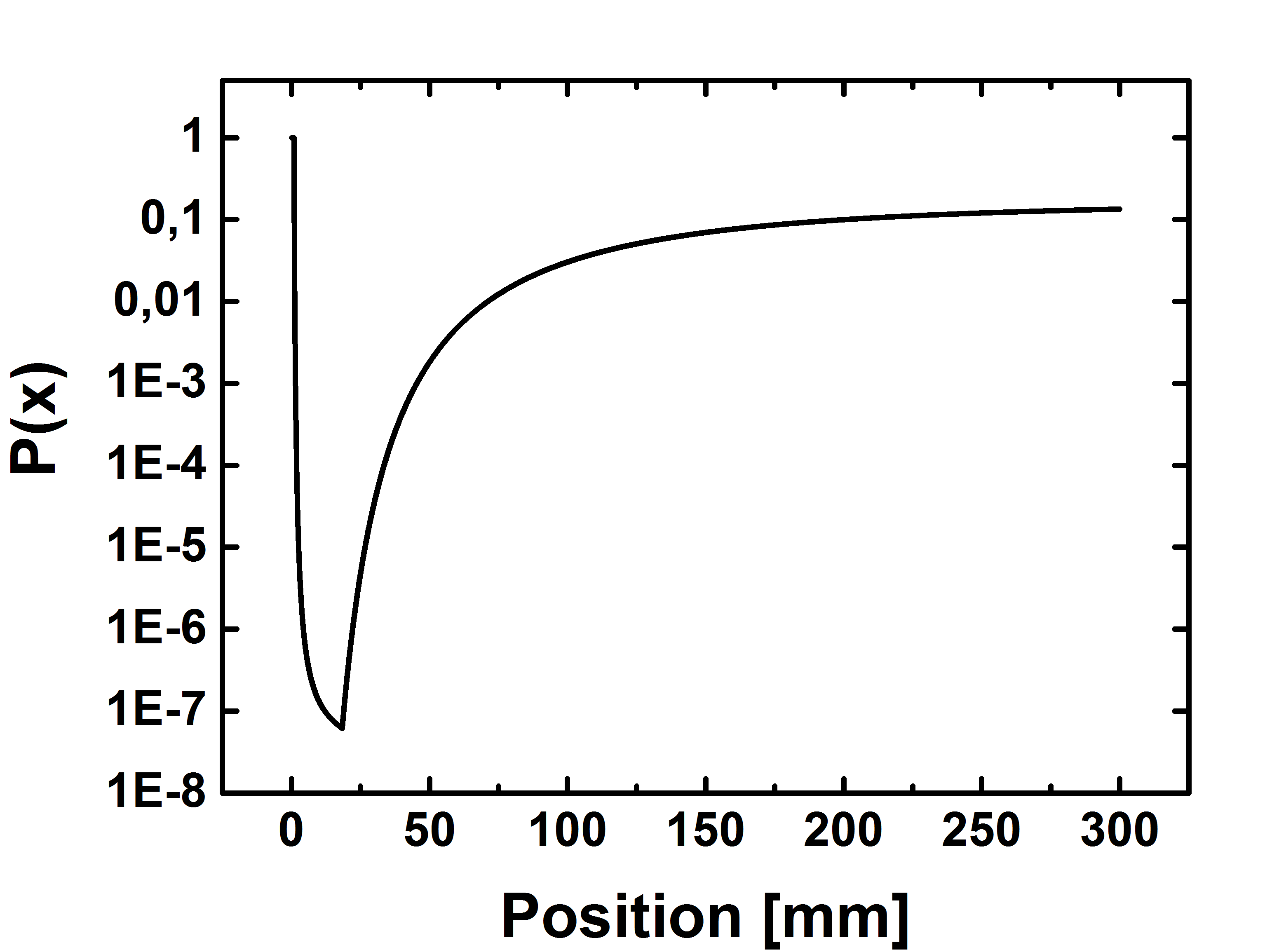}
\includegraphics [scale=0.058] {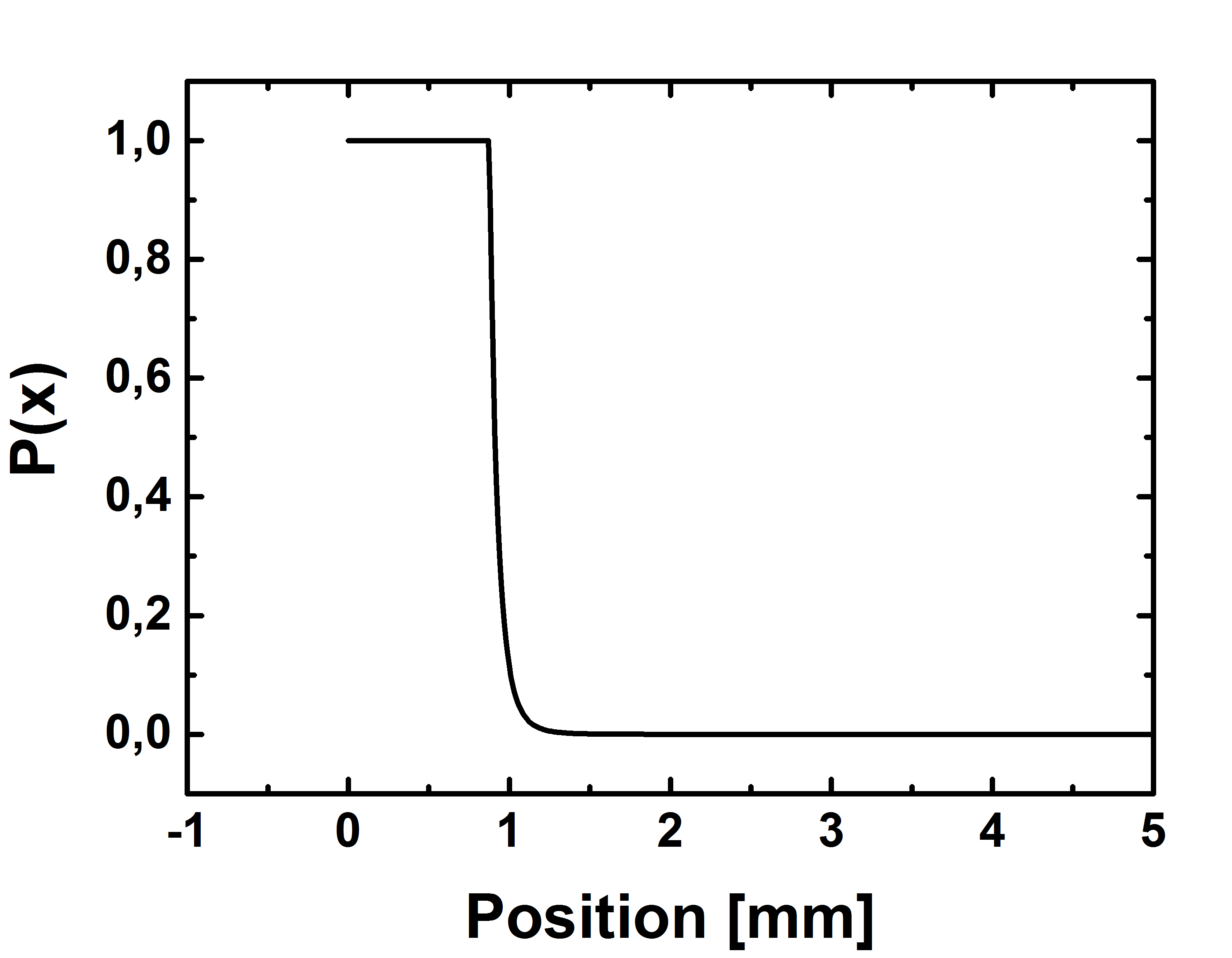}
\caption{(left) The probability of annihilation gamma quantum to pass through the collimator as a function of the position along the stripe ($x>0$) for $ s~=~1~mm $, $ r~=~100~mm $ and $ h~=~10~mm $ in logarithmic scale. (right) Zoom of the distribution for $0<x<5~mm$ in linear scale.}
\end{center}
\end{figure}

\newpage
\section{Conclusions}

\indent In order to design the optimal collimator system for the 24 module J-PET prototype, first we have studied a beam profile of the annihilation quanta emitted from the $^{22}$Na or $^{68}$Ge sources installed in the collimator with 1.3~mm wide slit and the two module detector. Three methods of the beam profile determination were presented. First method allowed for the 'wider' possible beam size estimation, second method contains the contribution from the gamma quanta that do not travel perpendicularly to the scintillator and the third method gives the estimation only for gamma quanta that travels perpendicularly to scintillator. Results obtained with all the methods are consistent within statistical uncertainties and show that the beam profile has a Gaussian shape with the FWHM equal to about 1~mm. The obtained result proves that the way of the collimation is suitable for the determination of the library of model signals required for the hit-time and hit-position determination when using reconstruction methods described in references \cite{10,12,13}. 

\indent For the 24 module detector not only width of the beam profile is important but also weight and size of collimator system, therefore to choose optimal solution simulations of the beam profile for different sizes of collimator were performed. What can be seen form the presented calculations is that the beam profile depends much more on the $r$ value than on $h$ (which was expected from purely geometrical considerations). Taking into account the weight and beam profile we have decided to construct two cylindrical collimators with dimensions: $ r~=~100~mm $, $ h~=~30~mm $ (Fig.~8) and $ r~=~100~mm $, $ h~=~10~mm $ (Fig.~9). For the latter we observe the increase of the probability of gamma quantum to pass through the collimator for $x>50~mm$, however this background could be rejected taking into account the time difference registered at both ends of the scintillator.

\section{Acknowledgements}

We acknowledge technical support by A.~Heczko, M.~Kajetanowicz, W.~Migdał and the financial support by the Polish National Center for Research and Development through grant INNOTECH-K1/1N1/64/159174/\\NCBR/12 and through LIDER grant 274/L-6/14/NCBR/2015, and the Foundation for Polish Science through MPD programme, and the EU and MSHE Grant No. POIG.02.03.00 - 161 00-013/09, and Marian Smoluchowski Krakow Research Consortium "Matter-Energy-Future".

%

\end{document}